\documentclass[twocolumn,showpacs,preprintnumbers,amsmath,amssymb,superscriptaddress,graphicx,graphics,pstricks]{revtex4}
\usepackage{amsmath}
\usepackage{graphicx}
\usepackage[latin1]{inputenc}
\usepackage{pstricks}
\usepackage{pst-node}
\usepackage{pst-coil}
\usepackage{pst-grad}
\usepackage{multido}
\usepackage{dcolumn}
\usepackage{bm}

\begin{document}
\title{Sommerfeld's image method in the calculation of van der Waals forces}

\author{Reinaldo de Melo e Souza}
\author{W.J.M. Kort-Kamp}
\author{C. Sigaud}
\author{C. Farina}
\affiliation{Instituto de Fisica, UFRJ, CP 68528, Rio de Janeiro,
RJ, 21941-972, Brasil}

\begin{abstract}
We show how the image method can be used together with a recent method developed
by C. Eberlein and R. Zietal to obtain the dispersive van der Waals interaction between an atom and a perfectly conducting surface
of arbitrary shape. We discuss in detail the case of an atom and a semi-infinite conducting plane.
In order to employ the above procedure to this problem it is necessary to use the ingenious image method introduced by Sommerfeld
more than one century ago, which is a generalization of the standard procedure. Finally, we briefly discuss other interesting situations that can also be treated by the joint use of Sommerfeld's image technique and Eberlein-Zietal method.

\end{abstract}

\maketitle

\section{Introduction}

Recent technological advances are responsible for the increasing number of experiments in the micro  and nano scales.
In such scales, dispersive forces become very important and, consequently, much
effort has been made in better understanding the role played by these forces. Many  papers have been dedicated recently
to the problem of atoms interacting with different surfaces, see \cite{McCauley2011}$^{-}$\cite{MendesEtAl-2008}
and references therein, to mention just a few.

In this paper we show how to employ a recent method developed by
C. Eberlein and R. Zietal \cite{Eberlein2007} in connection with the image technique largely used in electrostatic problems
 in order to obtain analytically expressions for the non-retarded dispersive interaction energy between an atom and a perfectly
 conducting surface. Particularly, we  explore this connection to obtain the
 dispersive van der Waals interaction between an atom and a semi-infinite conducting plane. Although one does not expect that the image method can be applied to this problem, an ingenious extension of the image method, provided by A. Sommerfeld in the late nineteenth century,
enables us to treat this and other interesting non-trivial systems by such a method.

This paper is organized as follows: in section 2 we briefly review
Eberlein-Zietal procedure and show how the image method can be used with it to yield
the non-retarded dispersive interaction between an atom and a perfectly conducting surface. In section
3 we show how to apply Sommerfeld's image method to obtain the electrostatic potential in the non-trivial system
constituted by a point charge in the presence of a semi-infinite conducting plane.
Since Sommerfeld's work is largely forgotten, we dedicate ourselves, in this section,  to present a pedagogical
exposition of Sommerfeld's main ideas. In section 4 we show how the methods introduced in  sections 2 and 3 can be used together to yield the quantum (non-retarded) dispersive interaction between an atom and a semi-infinite conducting plane, a result previously computed by other method by Eberlein and Zietal \cite{Eberlein2007}. In section 5 we discuss other geometries where Sommerfeld's image technique is also useful, giving special attention to the atom-disk system and the system constituted by an atom and an infinite plane with a circular aperture. Section 6 is left for the final remarks.

\section{Image method and the van der Waals force}

The use of the image method has rendered fruitful results in electrostatics, hydrodynamics, heat equation,
wave equation, etc. \cite{Macdonald1894}$^{-}$\cite{Hadamard1903}.
A recent method developed by C.Eberlein and
R.Zietal \cite{Eberlein2007} brings the possibility of a very simple and natural way of employing the
image method for the calculation of the van der Waals dispersive interaction of an atom and a perfectly conducting surface of arbitrary shape.
In this section we  briefly state this method and illustrate it with a very simple and well-known system, namely, an atom near an infinite conducting plane.

Let a neutral atom be at position $\mathbf{r}_0$ in the presence of an arbitrary
perfectly conducting surface $S$. In the non-retarded regime, i.e., for distances
much shorter than the dominant atomic transition wavelength, the dispersive interaction energy can be written in the form
 \cite{Eberlein2007}
\begin{equation}
 E_{int} = \frac{1}{2\varepsilon_0}\sum\limits_{i=1}^{3}\langle {\hat d}_i^2\rangle\partial_i\partial_i' G_H(\mathbf{r},\mathbf{r}')
 \bigg|_{\mathbf{r}=\mathbf{r}'=\mathbf{r}_0} \, , \label{eberlein}
\end{equation}
where ${\hat d}_i$ is the $i$ component of the atomic dipole operator ${\bf d}$ and $\langle\cdot\!\cdot\!\cdot\rangle$ means quantum expectation value taken with the atom in its ground state. The above expression is valid for any orthonormal coordinate system and for atoms with  no permanent dipole moment. Both conditions can be easily abandoned with only minor changes in the formula. For grounded conductors, the function $G_H$ appearing in equation (\ref{eberlein})
 satisfies the equation
\begin{equation}
		\nabla^2 G_H ({\bf r}, {\bf r}^{\,\prime})=0 \, , \label{GH}
\end{equation}
submitted to the boundary condition
\begin{equation}
	\left[\frac{1}{4\pi|\mathbf{r}-\mathbf{r}'|} + G_H(\mathbf{r},\mathbf{r}')\right]_{\mathbf{r}\in S}=0 \label{CCGH} \, .
\end{equation}
For non-grounded conductors, the boundary condition must be modified.
Previous equations are very similar to those satisfied by the potential $V_{i}(\mathbf{r})$ generated by
the image charges in the electrostatic problem of a charge $q$ at position ${\bf r}^{\,\prime}$ in the presence of a perfectly conducting grounded
surface $\mathcal{S}$, if the geometry accepts an image treatment. $V_{i}$ satisfies Laplace equation
and a boundary condition that differs from (\ref{CCGH}) only by a constant factor. We are thus allowed to make the
identification
\begin{equation}
G_H(\mathbf{r},\mathbf{r}') = \frac{\varepsilon_0 V_i({\bf r})}{q} \, . \label{ghim}
\end{equation}
Therefore,  to find the dispersive van der Waals interaction between an atom
and an arbitrary perfectly conducting surface all one ultimately needs is to solve an electrostatic problem. Afterwards,
by employing formula (\ref{eberlein}), one can obtain the desired quantum dispersive van der Waals interaction in a straightforward way.
This is not surprising since in the short distance regime (non-retarded regime) the retardation  of the fields can be neglected, allowing us to remain in the static regime. For distances of the order of the wavelength of atomic transitions, where retardation effects become important,
it is also possible to change from a quantum problem to a classical one, but it demands to solve an electrodynamic problem,
dealing with Helmholtz equation instead of Laplace equation. This is the essence of the scattering method, see
\cite{Reynaud2006} and references therein. For pedagogical reasons, we close this section by illustrating the procedure just described in a very elementary situation. Let the surface $S$ be an infinite conducting plane located at $y=0$. The image method
corresponding to this geometry is trivial and leads us, with the aid of (\ref{ghim}), to
\begin{equation}\label{GHExplicito}
G_H(\mathbf{r},\mathbf{r}') = - \frac{1}{4\pi\sqrt{(x - x^{\,\prime})^2 + (y + y^{\,\prime})^2
+ (z - z^{\,\prime})^2}} \, .
\end{equation}
Substituting this solution into eq. (\ref{eberlein}) we obtain the
well-known result\cite{Cohen1979}
\begin{equation}
	E_{0} = -\frac{\langle d_x^2\rangle+2\langle d_y^2\rangle+\langle d_z^2\rangle}{64\pi\epsilon_0|y_0|^3} \, ,
\label{eap}
\end{equation}
This example shows the simplicity of the image method in dealing with dispersive
interactions. The pitfall seems to be the small number of geometries compatible with this kind of treatment.
Nevertheless, that number was fortunately increased with the ingenious extension developed by Arnold
Sommerfeld in the end of the 19th century, as we shall see along this work.
%
%

\section{Sommerfeld's method for a  charge and a semi-infinite plane}

\begin{figure}
\begin{center}
\newpsobject{showgrid}{psgrid}{subgriddiv=1,griddots=10,gridlabels=6pt}
	\scalebox{0.60}
{
\begin{pspicture}(0,-5.195)(11.64,5.235)
\psline[linewidth=0.02cm,arrowsize=0.15291667cm 2.0,arrowlength=1.4,arrowinset=0.4]{->}(2.62,-1.845)(2.62,4.155)
\psline[linewidth=0.02cm,arrowsize=0.15291667cm 2.0,arrowlength=1.4,arrowinset=0.4]{->}(0.0,-1.105)(12.82,-1.105)
\psline[linewidth=0.04cm](2.62,-1.105)(2.62,1.075)
\psline[linewidth=0.04cm,arrowsize=0.15291667cm 2.0,arrowlength=1.4,arrowinset=0.4]{<-}(2.62,0.895)(2.62,3.315)
\psline[linewidth=0.04cm](2.62,3.315)(4.3,3.315)
\psline[linewidth=0.04cm,arrowsize=0.15291667cm 2.0,arrowlength=1.4,arrowinset=0.4]{<-}(4.18,3.315)(5.78,3.315)
\psline[linewidth=0.04cm](5.78,3.315)(5.78,2.215)
\psline[linewidth=0.04cm,arrowsize=0.15291667cm 2.0,arrowlength=1.4,arrowinset=0.4]{<-}(5.78,2.315)(5.78,1.195)
\psline[linewidth=0.04cm](6.158,1.315)(6.158,2.135)
\psline[linewidth=0.04cm,arrowsize=0.15291667cm 2.0,arrowlength=1.4,arrowinset=0.4]{<-}(6.158,2.015)(6.158,3.315)
\psline[linewidth=0.02cm,linestyle=dashed,dash=0.16cm 0.16cm](5.94,4.035)(5.94,1.355)
\psline[linewidth=0.02cm,linestyle=dotted,dash=0.16cm 0.16cm](5.94,1.355)(5.94,-2.905)
\psline[linewidth=0.02cm,linestyle=dashed,dash=0.16cm 0.16cm](5.94,-2.905)(5.94,-5.185)
\rput{-180.0}(11.92,2.59){\psarc[linewidth=0.04](5.96,1.295){0.2}{30.068583}{191.30994}}
\psdots[dotsize=0.1](5.94,1.355)
\psline[linewidth=0.02cm,linestyle=dotted,dash=0.16cm 0.16cm](1.74,1.355)(5.9,1.355)
\psline[linewidth=0.04cm](5.74,-2.685)(5.74,-3.785)
\psline[linewidth=0.04cm,arrowsize=0.15291667cm 2.0,arrowlength=1.4,arrowinset=0.4]{<-}(5.74,-3.685)(5.74,-4.705)
\psline[linewidth=0.04cm](6.118,-4.685)(6.118,-3.865)
\psline[linewidth=0.04cm,arrowsize=0.15291667cm 2.0,arrowlength=1.4,arrowinset=0.4]{<-}(6.118,-3.985)(6.118,-2.685)
\rput{-35.0}(2.6784291,2.8905418){\psarc[linewidth=0.04](5.9230285,-2.8021753){0.18766797}{30.068583}{191.30994}}
\psdots[dotsize=0.12](5.94,-2.905)
\psline[linewidth=0.02cm,linestyle=dotted,dash=0.16cm 0.16cm](1.74,-2.925)(5.9,-2.925)
\psline[linewidth=0.04cm,arrowsize=0.15291667cm 2.0,arrowlength=1.4,arrowinset=0.4]{<-}(6.14,3.315)(8.62,3.315)
\psline[linewidth=0.04cm,arrowsize=0.15291667cm 2.0,arrowlength=1.4,arrowinset=0.4]{<-}(9.52,3.315)(11.5,3.315)
\psline[linewidth=0.04cm](11.5,3.335)(11.5,-0.685)
\psline[linewidth=0.04cm](2.6,-4.645)(2.62,-0.965)
\usefont{T1}{ptm}{m}{n}
\rput(12.73,-1.52){\Large {Re $\alpha$}}
\usefont{T1}{ptm}{m}{n}
\rput(1.95,4.24){\Large{Im $\alpha$}}
\usefont{T1}{ptm}{m}{n}
\rput(2.47,-1.36){\Large{0}}
\usefont{T1}{ptm}{m}{n}
\rput(11.12,-1.46){\Large{$4\pi$}}
\usefont{T1}{ptm}{m}{n}
\rput(6.28,-1.44){\Large$\varphi'$}
\usefont{T1}{ptm}{m}{n}
\rput(2.26,1.6){\Large$\gamma$}
\usefont{T1}{ptm}{m}{n}
\rput(2.2,-3.14){\Large$-\gamma$}
\psline[linewidth=0.04cm,arrowsize=0.15291667cm 2.0,arrowlength=1.4,arrowinset=0.4]{->}(6.08,-4.705)(8.66,-4.705)
\psline[linewidth=0.04cm](9.52,-4.685)(11.52,-4.685)
\psline[linewidth=0.04cm,arrowsize=0.15291667cm 2.0,arrowlength=1.4,arrowinset=0.4]{->}(2.58,-4.665)(4.26,-4.665)
\psline[linewidth=0.04cm](4.14,-4.665)(5.74,-4.665)
\usefont{T1}{ptm}{m}{n}
\rput(3.65,5.04){\color{white} Oi}
\psdots[dotsize=0.1](3.98,-1.105)
\usefont{T1}{ptm}{m}{n}
\rput(3.99,-1.46){\Large$\varphi$}
\psline[linewidth=0.04cm,arrowsize=0.15291667cm 2.0,arrowlength=1.4,arrowinset=0.4]{<-}(11.5,-0.625)(11.5,-4.665)
\psline[linewidth=0.04cm](8.68,3.315)(8.68,2.215)
\psline[linewidth=0.04cm,arrowsize=0.15291667cm 2.0,arrowlength=1.4,arrowinset=0.4]{<-}(8.68,2.315)(8.68,1.195)
\psline[linewidth=0.04cm](9.058,1.315)(9.058,2.135)
\psline[linewidth=0.04cm,arrowsize=0.15291667cm 2.0,arrowlength=1.4,arrowinset=0.4]{<-}(9.058,2.015)(9.058,3.315)
\psline[linewidth=0.02cm,linestyle=dashed,dash=0.16cm 0.16cm](8.84,4.035)(8.84,1.355)
\psline[linewidth=0.02cm,linestyle=dotted,dash=0.16cm 0.16cm](8.84,1.355)(8.84,-2.905)
\psline[linewidth=0.02cm,linestyle=dashed,dash=0.16cm 0.16cm](8.84,-2.905)(8.84,-5.185)
\rput{-180.0}(17.72,2.59){\psarc[linewidth=0.04](8.86,1.295){0.2}{30.068583}{191.30994}}
\psdots[dotsize=0.1](8.84,1.355)
\psline[linewidth=0.02cm,linestyle=dotted,dash=0.16cm 0.16cm](4.64,1.355)(8.8,1.355)
\psline[linewidth=0.04cm](8.64,-2.685)(8.64,-3.785)
\psline[linewidth=0.04cm,arrowsize=0.15291667cm 2.0,arrowlength=1.4,arrowinset=0.4]{<-}(8.64,-3.685)(8.64,-4.705)
\psline[linewidth=0.04cm](9.018,-4.685)(9.018,-3.865)
\psline[linewidth=0.04cm,arrowsize=0.15291667cm 2.0,arrowlength=1.4,arrowinset=0.4]{<-}(9.018,-3.985)(9.018,-2.685)
\rput{-35.0}(3.2028883,4.5539136){\psarc[linewidth=0.04](8.823029,-2.8021753){0.18766797}{30.068583}{191.30994}}
\psdots[dotsize=0.12](8.84,-2.905)
\psline[linewidth=0.02cm,linestyle=dotted,dash=0.16cm 0.16cm](4.64,-2.925)(8.8,-2.925)
\psline[linewidth=0.04cm](9.06,3.275)(9.6,3.295)
\psline[linewidth=0.04cm,arrowsize=0.15291667cm 2.0,arrowlength=1.4,arrowinset=0.4]{->}(9.02,-4.685)(9.78,-4.685)
\usefont{T1}{ptm}{m}{n}
\rput(8.68,-1.46){\Large$\varphi'+2\pi$}
\end{pspicture}
}
\end{center}
\caption{Choice of the contour. The cuts are represented by the dashed lines.}
\label{Circuit}
\end{figure}
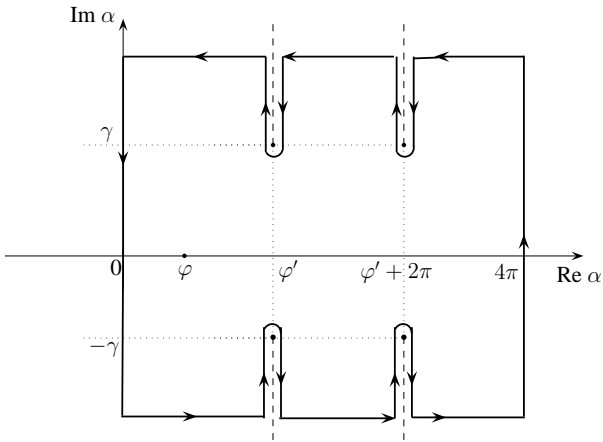

To begin with, consider a charge $q$ in the presence of a semi-infinite perfectly conducting
plane, taken as $y=0$, $x\geq 0$. As discussed in the last section, to find the electrostatic potential at every point of the space (where it is defined) for this problem is all we need to obtain the van der Waals interaction between an atom and a semi-infinite conducting plane.
The appropriate system of coordinates for this problem is the cylindrical set $(r,\varphi,z)$,
since with this choice the equation of the semi-plane is only $\varphi=0,2\pi$.
Naively, one may assume that image method is unsuited for this geometry once all points of
space seem to be already used, leaving no room for image charges. Sommerfeld's procedure
\cite{Sommerfeld1896}$^{,}$\cite{Davis1971} is based on creating somehow space to allocate the image charges.
In his construction, to come back to the point of departure a rotation
of $4\pi$ is required. In other words, the points $(r,\varphi,z)$ and $(r,\varphi+2\pi,z)$ are
no longer the same. In this way, Sommerfeld revived the riemannian sheets of complex analysis,
by constructing a two-fold space - points with $0\leq\varphi<2\pi$ belong to the ordinary space
while those with $2\pi\leq\varphi<4\pi$ constitute the auxiliary space, where we will be able to locate the necessary  image charges.
Note that with this construction we create a discontinuity at the conducting surface. Indeed,
when we approach the semi-plane from above in the ordinary space we have $\varphi\rightarrow 0$, while
when we approach from below, we have $\varphi\rightarrow 2\pi$.
Let us consider the charge $q$ at position $(r',\varphi',z')$.  The potential generated by this charge
in the double space cannot be the familiar $1/R$, with $R=|\mathbf{r}-\mathbf{r}'|$.
This is not surprising since the potential created
by a point charge depends on the geometry of the space - recall that in 2 dimensions the coulomb potential
is not any more proportional to 1 over the distance from the point of space and the position of the charge, but it is given by a logarithm of such a distance over a reference distance. The potential $1/R$ presents a symmetry
by changing $\varphi$ by $\varphi+2\pi$, so its laplacian furnishes two Dirac delta functions, one
with singularity at $(r',\varphi',z')$ and another with singularity at $(r',\varphi'+2\pi,z')$.
We are thus led to recognize in $1/R$ the superposition of the potential of two distinct charges, one
in the ordinary space and the other in the correspondent place of the auxiliary space. In order
to identify each term, we  employ Cauchy theorem, which states that
\begin{equation}
	\label{cauchy}
	\frac{1}{R(w)}=\frac{1}{2\pi i}\oint_C \frac{R^{-1}(w')}{w'-w}dw \, ,
\end{equation}
where $C$ is any closed contour in the complex plane, provided that it encloses
the pole $w'=w$ and $R^{-1}(w')$ is analytical in the interior of and along $C$. We must chose a convenient variable $w$. Since $\varphi$ is the variable which operates the passage from the ordinary to the auxiliary space, the smart choice is
\begin{equation}
	w=e^{i\varphi/2} \, , \label{w}
\end{equation}
which has a period of $4\pi$, as the double space. Accordingly, we set $w'=e^{i\alpha/2}$ which leaves equation (\ref{cauchy}) in the form
\begin{equation}
	\label{cauchyalpha}
	\frac{1}{R(\varphi)}=\frac{1}{4\pi}\oint_C \frac{1}{R(\alpha)} \frac{e^{i\alpha/2}}{e^{i\alpha/2}-e^{i\varphi/2}} d\alpha\, .
\end{equation}
Now, all we have to do is to choose the contour $C$. The distance between points $(r,\varphi,z)$ and
$(r^{\,\prime},\varphi^{\,\prime},z^{\,\prime})$ is given by $R^2=r^2+r'^2-2rr'\cos(\varphi-\varphi')+(z-z')^2$.
$R({\alpha})$ is obtained from $R$ just changing $\varphi$ by $\alpha$. The presence of the square root in $R^{-1}(\alpha)$ generates a branch cut. The branch points are determined by the equation $R(\alpha)=0$, whose solutions are $\alpha=\varphi'+2m\pi\pm i\gamma$,
$ m\in\mathrm{Z}$, with the definition
\begin{equation}
	\cos(i\gamma) = \frac{r^2+r'^2+(z-z')^2}{2rr'} \label{gammadef}  \, .
\end{equation}
Our contour must avoid the cut generated on these branch points. One possibility is sketched in the Figure
\ref{Circuit}.
Now, we shall perform the integration (\ref{cauchyalpha}) in this contour. Due to the periodicity of the
integrand, the integrations at vertical lines Re$\,\alpha=0$ and Re$\,\alpha=4\pi$ cancel out. The contributions
of the horizontal paths vanish in the limit where the height of the circuit in Figure \ref{Circuit} is taken to
 infinite. This is evident once $R_{\alpha}^{-1}\rightarrow 0$ in this limit, leaving the integrand of
(\ref{cauchyalpha}) null there. We are left with the integrals around the cuts. Let us call $A_0$ the path
around the branch points $\varphi'\pm i\gamma$ and $A_1$ the one around the branch points
$\varphi'+2\pi+\pm i\gamma$. Therefore, the potential $1/R$ may be written as
\begin{equation}
	\label{potespdup}
		\frac{1}{R}=\frac{1}{4\pi }\int\limits_{\, A_0} \frac{R_{\alpha}^{-1}}{1-e^{i(\varphi-\alpha)/2}} d\alpha +\frac{1}{4\pi }\int\limits_{\, A_1} \frac{R_{\alpha}^{-1}}{1-e^{i(\varphi-\alpha)/2}} d\alpha\, .
\end{equation}
We achieved a separation of the potential $1/R$ in two terms. Sommerfeld showed that the first
{\it (i)} is uniquely defined, finite and continuous at all points of the double space, except at $(r',\varphi', z')$. This means, particularly,
that it is finite at $(r',\varphi'+2\pi,z')$;  {\it (ii)} has a null laplacian  at all points except at $(r',\varphi', z')$ and at the surface of the conductor;
  {\it (iii)} vanishes at infinity and
  {\it (iv)} is bivalent in the ordinary space, with a separated branch to each copy of the double space.
Therefore, including the necessary constants, we realize that the potential of a single charge at position
$(r',\varphi',z')$ in the double space is identified as
\begin{equation}
	V(r,\varphi,z)=\frac{q}{16\pi^2\varepsilon_0}\int \limits_{\, A_0} \frac{R_{\alpha}^{-1}}{1-e^{i(\varphi-\alpha)/2}} d\alpha \, . \label{vparcial}
\end{equation}
Analogously, the term containing the integration along the path $A_1$ can be identified as the potential in the double space
created by a charge  at position $(r',\varphi' + 2\pi,z')$. To evaluate the integral in the previous equation,
we shall first observe that the contour $A_0$ has two paths, one
around each branch point, and each path contains three parts, two verticals and a semi-circle around
the branch point. In the limit where the radius of this semi-circle goes to zero, the integral along the
semi-circle vanishes. The cut prevents the contributions of the two vertical lines to cancel each other. It can be shown
from complex analysis that they give the same result. In order to calculate it,
we parametrize the vertical line around the branch point $\alpha=\varphi'+i\gamma$ by $\alpha=\varphi'+i\beta$,
and the one around the branch point below, $\alpha=\varphi'-i\gamma$, by $\alpha=\varphi'-i\beta$. This
way, both integrals run with $\beta$ varying from $\gamma$ to $\infty$, and
a factor 2 must be included to account for both of them. Hence, equation (\ref{vparcial}) becomes
\begin{eqnarray}
&&V(\rho,\theta,\varphi)=\frac{q}{8\pi^2\varepsilon_0\sqrt{rr'}}\times\nonumber\\
&&\int\limits_{\gamma}^{\infty} \frac{(\cosh\beta-\cosh\gamma)^{-1/2}\sinh(\beta/2)}{\cosh(\beta/2)-\cos\left(\frac{\varphi-\varphi'}{2}\right)}\,
d\beta \, . \label{intao}
\end{eqnarray}
This integral can be recast with a convenient change of variables. Defining
\begin{equation}
	\xi:=\cosh(\beta/2) \, , \sigma:=\cosh(\gamma/2) \, , \tau:=\cos\left(\frac{\varphi-\varphi'}{2}\right) \, ,
	\label{xi}
\end{equation}
we rewrite integral (\ref{intao}) as
\begin{eqnarray}
	 V(r,\varphi,z)&=&\frac{q}{8\pi^2\varepsilon_0\sqrt{2rr'}}\int\limits_{\sigma}^{\infty}\frac{d\xi}{(\xi^2-\sigma^2)^{1/2}(\xi-\tau)}\nonumber\\
&=&\frac{q}{2\pi^2\varepsilon_0R}\tan^{-1}\left[\frac{(\sigma+\tau)}{(\sigma-\tau)}\right]^{1/2} \, . \label{v2pronto}
\end{eqnarray}
Now that we know the potential of a point charge in the double space, we may proceed with
the image technique to find the electrostatic potential of a charge $q$ in the presence
of a semi-infinite plane. It is easily seen that if we put an image charge $-q$ at
$(r',4\pi-\varphi',z')$ we have the boundary conditions satisfied. Indeed, the potential
of the system constituted by these two charges is
\begin{eqnarray}
		 V_{sp}(r,\varphi,z)&=&\frac{q}{2\pi^2\varepsilon_0R}\tan^{-1}
 \left[\frac{(\sigma+\tau)}{(\sigma-\tau)}\right]^{1/2}+\nonumber\\
&-&\frac{q}{2\pi^2\varepsilon_0R_i}\tan^{-1}
 \left[\frac{(\sigma+\tau_i)}{(\sigma-\tau_i)}\right]^{1/2} \, , \label{potsp}
\end{eqnarray}
where $R_i$ and $\tau_i$ are obtained from $R$ and $\tau$ just changing $\varphi'$ by $4\pi-\varphi'$ and $\sigma$
is the same for both charges since it doesn't depend on $\varphi'$. At the conducting surface, characterized
by $\varphi=0,2\pi$, we have $R=R_i$ and $\tau=\tau_i$. Therefore, $V_{sp}$ vanishes at the
conducting semi-infinite plane. Observe that for $\varphi=\pi$, we have $V_{sp}\neq 0$, as expected.
Note, also, that the position occupied by the image charge is the correspondent point in the
auxiliary space of the mirror image of the charge. Henceforth, we recognize the potential (\ref{potsp}) as
the potential of the configuration formed by a charge and a semi-infinite plane.

\section{An atom and a semi-infinite plane}

To employ relation (\ref{ghim}) we must first determine the potential $V_i$ created by the image charge in
the charge-semi-infinite plane configuration. This is easily done once we know the complete potential,
given by (\ref{potsp}),
\begin{equation}
	V_i(r,\varphi,z) = V_{sp}(r,\varphi,z) - \frac{q}{4\pi\varepsilon_0R} \, .
\end{equation}
Substituting this into equation (\ref{ghim}) we obtain
\begin{eqnarray}
	 G_H(\mathbf{r},\mathbf{r}')&=&-\frac{1}{2\pi^2R}\tan^{-1}
 \left[\frac{(\sigma-\tau)}{(\sigma+\tau)}\right]^{1/2}+\nonumber\\
&-&\frac{1}{2\pi^2R_i}\tan^{-1}
 \left[\frac{(\sigma+\tau_i)}{(\sigma-\tau_i)}\right]^{1/2} \, . \label{ghsp}
\end{eqnarray}
Applying (\ref{eberlein}), we obtain the desired van der Waals interaction between an atom at $\mathbf{r}_0$ and a semi-infinite plane,
\begin{equation}
		E_{sp}=\frac{-1}{4\pi\varepsilon_0} [A\langle d_r^2\rangle+B\langle d_{\varphi}^2\rangle+C\langle d_z^2\rangle]\, ,
	\end{equation}
with 	
\begin{eqnarray}
		A&=&\frac{5}{48\pi r_0^3}+\dfrac{\cos\varphi_0}{16\pi r_0^3\sin^2\varphi_0}+\frac{|\pi-\varphi_0|(1+\sin^2\varphi_0)}{16\pi r_0^3\sin^3\varphi_0} \label{a} \nonumber \\
			B&=&-\frac{1}{48\pi r_0^3}+\dfrac{\cos\varphi_0}{8\pi r_0^3\sin^2\varphi_0}+\frac{|\pi-\varphi_0|(1+\cos^2\varphi_0)}{16\pi r_0^3\sin^3\varphi_0} \nonumber \\
		C&=&\frac{1}{24\pi r_0^3}+\dfrac{\cos\varphi_0}{16\pi r_0^3\sin^2\varphi_0}+\frac{|\pi-\varphi_0|}{16\pi r_0^3\sin^3\varphi_0} \, . \label{c}
	\end{eqnarray}
Note that, by symmetry arguments, the interaction energy is independent of $z_0$. This
problem was analysed by C. Eberlein and R.Zietal \cite{Eberlein2007} who solved it
by employing properties of Bessel functions. It is possible to show that the $G_H$ obtained
there for this geometry is equivalent to our equation (\ref{ghsp}). It
is instructive to analyse our results in the limit where the atom is very close to the conducting
surface and far from the edge. To do so, we must take the limits $r_0\rightarrow\infty$,
$\varphi_0\rightarrow 0$, keeping the product $y_0=r_0\sin\varphi_0$ constant. Doing this,
we get $A\rightarrow  1/(16 y_0^3)\, , B\rightarrow 1/(8 y_0^3)$ and
$C\rightarrow 1/(16 y_0^3)$, which allows us to recast the interaction energy  in the form
\begin{equation}
		E_{sp}\approx \frac{-1}{64\pi\varepsilon_0y_0^3} [\langle d_x^2\rangle+2\langle d_{y}^2\rangle+\langle d_z^2\rangle] \, ,
\end{equation}
As expected, last result coincides with that obtained for an atom in the presence of an infinite plane, see equation (\ref{eap}).

\section{Other non-trivial applications}

As already mentioned, analytical expressions of dispersive interactions
for different geometries are important in a variety of situations. Hence, the
natural question one may pose is whether there are other geometries which admit a treatment by the image method.
The purpose of this section is to answer this question positively.
In the solution for the semi-plane, we employed a two-fold space. Some times, we need to deal
with more spaces. As an example, think of an atom in the presence of a wedge. If the angle
of the wedge is $\pi/m$, with $m$ being a positive integer, we may employ the standard image technique to
solve the electrostatic problem and then apply Eberlein-Zietal formula. On the other hand, let
the angle be $n\pi/m$, with $n,m$ positive integers. The standard image procedure  yields images on
the region inside the wedge, leading to the failure of the method. In spite of this, we may
approach the problem through Sommerfeld's generalization. Consider, without loss of generality,
an irreducible fraction $n/m$. To apply the formalism of the last section we will have to deal with
$n$ spaces. This makes the calculations a little harder, but the ideas are the same. To give a sample,
the fundamental modification is the variable we choose in Cauchy's theorem (\ref{cauchy}). Since
the $n$-fold space has a period $2\pi n$, instead of (\ref{w}), we must choose
\begin{equation}
	w=e^{i\varphi/n} \, .
\end{equation}
As a consequence, we must substitute equation (\ref{cauchyalpha}) by
\begin{equation}
	\frac{1}{R(\varphi)}=\frac{1}{2\pi n}\oint_C R^{-1}(\alpha) \frac{e^{i\alpha/n}}{e^{i\alpha/n}-e^{i\varphi/n}} d\alpha\, .
\end{equation}
This potential is the superposition of the potential of $n$ point charges, located at
$(r',\varphi'+i2\pi,z'), (i=0,..,n-1)$.
 The decomposition
of the potential $1/R$ into $n$ contributions is done in a similar fashion. We must
enlarge the circuit of Figure \ref{Circuit}, by changing the vertical line at Re$\,\alpha=4\pi$ by
the vertical line at Re$\,\alpha=2\pi n$, so that after integration the vertical contributions
cancel out due to the periodicity of the integrand. This enlargement brings more cuts to the integral,
leading to the desired extra terms in (\ref{potespdup}), which becomes
\begin{equation}
		\frac{1}{R}=\sum\limits_{j=0}^{n-1}\frac{1}{2\pi n }\int\limits_{\, A_j} \frac{R_{\alpha}^{-1}}{1-e^{i(\varphi-\alpha)/2}} d\alpha \, .
\end{equation}
From the operational point of view, the only change from the two-fold space to
the $n$-fold one is the integration we must perform. For more details, see the original
memoir \cite{Sommerfeld1896}.

In the last paragraph we saw other possibilities of employing the image technique increasing
the number of folds of the space. We may find also examples that require only the two-fold space
treated before. To this end, we generally have to use different system of coordinates. It is
of foremost importance in the image method to describe the problem by the system of coordinates that
suits it. A relevant example is that of an atom interacting with an infinite plane with a circular aperture,
which leads to repulsion as shown by Levin et al.\cite{Levin2010}. In a recent
paper \cite{nos}, we treated this problem by Sommerfeld's image method and obtained an analytical result
for the corresponding interaction energy of this system. Our results agree with those previously obtained by Eberlein
and Zietal \cite{Eberlein2011}. We close this section by summing up our main results for a system constituted by an atom and an infinite conducting plane with a circular aperture and for the atom-disk system.

The interaction energy  between an atom and an infinite plane with a circular aperture is
found in a similar way as we did in the semi-plane system. For an atom on
the symmetry axis of the aperture, say the $z$-axis, and with major pollarizability
in the $z$ direction, our expression becomes very simple and it is given by
\begin{eqnarray}
E_{ph} &=&-\frac{\langle d_z^2\rangle}{64\varepsilon_0\pi z^3}\Biggl[1+\frac{2}{\pi}\sin^{-1}\left(\frac{z^2-a^2}{z^2+a^2}\right)+\nonumber\\
&-&\frac{4az(3a^4+8a^2z^2-3z^4)}{3\pi(a^2+z^2)^3}\Biggr] \, ,\label{apbexato}
\end{eqnarray}
where $a$ is the radius of the aperture. For small $z$ we get repulsion. We found an equilibrium point
at $z_{eq}\approx 0.74235 a$. For $z>z_{eq}$ the interaction is attractive so that the equilibrium is stable in the
 $z$ direction. However, it can be shown that for lateral displacements the equilibrium is unstable.
Atomic anisotropy is essential to get repulsion. With this method we may also treat the complementary geometry, namely, the atom-disk system.
The interaction energy when the atom is in the symmetry axis of the disk and with dominant pollarizability in this
direction is \cite{nos}
\begin{eqnarray}
	E_{disk}&=&-\frac{\langle d_z^2\rangle}{64\varepsilon_0\pi z^3}
\Biggl[1-\frac{2}{\pi}\sin^{-1}\left(\frac{z^2-a^2}{z^2+a^2}\right)+\nonumber\\
&+&\frac{4az(3a^4+4a^2z^2+9z^4)}{3\pi(a^2+z^2)^3}\Biggr]\! , \label{adexato}
\end{eqnarray}
where $a$ is now the radius of the disk. Expressions (\ref{apbexato}) and (\ref{adexato}) allow us to make a quantitative study
of the finite-size effect for those geometries. Other interesting feature is
the possibility of an exact calculation of the non-additivity effects. Indeed,
evaluating $F_{disk}=-\partial_z E_{disk}$ and summing it with $F_{ph}=-\partial_z E_{ph}$ we obtain
\begin{equation}
F_{disk}+F_{ph} = F_0 - \dfrac{\langle d_z^2\rangle a}{\varepsilon_0\pi^2} \dfrac{(z^2-a^2)z}{(z^2+a^2)^4} \, ,
\label{NAforce}
\end{equation}
where $F_0$ is the atom-infinite plane force.
We see that non-addivity vanishes for $a\rightarrow\infty$ and for $a\rightarrow 0$, as it should.
It is remarkable the existence of a distance for which the superposition holds.
In eq.(\ref{NAforce}) this distance is found to be $z=a$.
%

\section{Final remarks}
\!\!
The main purpose of this paper was to establish the image method as a powerful tool to calculate
non-retarded dispersive forces between an atom and a conducting surface. The recent procedure developed by
Eberlein and Zietal \cite{Eberlein2007} is remarkably and naturally linked to the electrostatic image method. As an example,
we discussed in detail the non-trivial system constituted by an atom and a semi-infinite conducting plane. This problem
reveals the importance of choosing an adequate system of coordinates, since
it is the angular variable $\varphi$ which sustains and operates the two-fold construction of the Sommerfeld's image method.
We briefly analysed in a similar way two other non-trivial systems, namely, an atom and an infinite conducting plate with
a circular aperture and the atom-disk system. Sommerfeld's image method is not systematic in the sense that for each different
geometry one has to work out the proper system of coordinates and develop the multi-fold
space. On one hand, this may be considered a pitfall but, on the other hand, it leaves room
for the creativity and imagination of the researcher when challenged to solve non-trivial problems.
Once the exact analytical dispersive interaction energy is obtained, a variety of questions may
be answered. In this connection, we would like to mention particularly the possibility
of studying finite size effects and, in some cases, non-additivity effects involving complemetary surfaces. In the latter case, the
complementary geometries of an atom-disk system and  the system constituted by an atom and a plane with a circular aperture
was studied in [\cite{nos}], where a surprising result was found, namely: that there exists a distance between the atom and
the center of the aperture for which the non-additivity effects vanish.
The authors believe this result may occur in other systems involving complementary surfaces.

%
%
%
\noindent
{\bf Acknowledgements}\\
The authors are indebted with P.A. Maia Neto, F.S.S. Rosa, F. Pinheiro, A.L.C. Rego and Marluce Faria for
valuable discussions. The authors  also  thank to CNPq and FAPERJ (brazilian agencies) for
partial financial support.


\end{document}